\documentclass[aps,prb,twocolumn,superscriptaddress,showpacs]{revtex4}

\usepackage{amsmath,amssymb}
\usepackage{bm}% bold math
\usepackage{graphicx}% complex graphics
\usepackage{float,afterpage,wrapfig}

\newcommand{\Tm}{\mathcal{T}}
\newcommand{\veps}{\varepsilon}

\begin{document}

\title{Finite-temperature conductance signatures of
quantum criticality in double quantum dots}

\author{Luis G.\ G.\ V.\ Dias da Silva}
\affiliation{Department of Physics and Astronomy and Nanoscale and
Quantum Phenomena Institute, Ohio University, Athens, Ohio
45701--2979}
\affiliation{Materials Science and Technology Division, Oak Ridge
National Laboratory, Oak Ridge, TN 37831, and Department of Physics
and Astronomy, University of Tennessee, Knoxville, TN 37996}

\author{Kevin Ingersent}
\affiliation{Department of Physics, University of Florida, P.O.\
Box 118440, Gainesville, Florida, 32611--8440}

\author{Nancy Sandler}
\affiliation{Department of Physics and Astronomy and Nanoscale and
Quantum Phenomena Institute, Ohio University, Athens, Ohio 45701--2979}

\author{Sergio E. Ulloa}
\affiliation{Department of Physics and Astronomy and Nanoscale and
Quantum Phenomena Institute, Ohio University, Athens, Ohio 45701--2979}

\date{\today}

\begin{abstract}
We study the linear conductance through a double-quantum-dot system
consisting of an interacting dot in its Kondo regime and an effectively
noninteracting dot, connected in parallel to metallic leads. Signatures
in the zero-bias conductance at temperatures $T>0$ mark a pair of quantum
($T=0$) phase transitions between a Kondo-screened many-body ground state
and non-Kondo ground states. Notably, the conductance features become more
prominent with increasing $T$, which enhances the experimental prospects
for accessing the quantum-critical region through tuning of gate voltages
in a single device.
\end{abstract}

\pacs{73.63.Kv, 73.43.Nq, 72.15.Qm}
% 73.43.Nq Quantum phase transitions
% 73.63.Kv  Quantum dots
% 72.15.Qm Scattering mechanisms and Kondo effect
\maketitle

Quantum phase transitions (QPTs) occur in the zero temperature
($T=0$) phase diagram of a system at points of nonanalyticity of
the ground-state energy. \cite{Sachdev:book,Vojta:1807:2006} QPTs
underlie many fascinating phenomena in strongly interacting
condensed matter, including the metal-insulator transition in
disordered systems, \cite{MIT} the destruction of
antiferromagnetism with doping in high-temperature superconductor
parent compounds, \cite{HTC} the magnetic-field-driven
superconducting-insulator transition in disordered
superconductors, \cite{SIT} and quantum Hall plateau transitions.
\cite{qH} Study of most of these QPTs is hindered by the need to
fabricate controlled series of samples at different
stoichiometries and/or disorder levels.

By contrast, it is increasingly apparent that systems of quantum
dots offer possibilities for exploring QPTs (strictly,
\textit{boundary} QPTs involving only a subset of the system
degrees of freedom) within a single sample. Advances in system
fabrication, precise characterization, and the near suppression of
dissipative and incoherent environments
\cite{Goldhaber-Gordon:156:1998} have enabled beautiful
experiments on multi-dot devices. \cite{multi-dot_expts} This leap
forward in experimental capability has also spurred much
theoretical activity, including several predictions of QPTs in
quantum dots in the Kondo regime. \cite{QPTsNRG} The feasibility
of realizing nontrivial many-body states has been confirmed by the
recent experimental demonstrations of a two-channel Kondo regime
\cite{Potok:167:2007} and of a singlet-triplet QPT.
\cite{Roch:633:2008}

This Letter predicts robust signatures of QPTs in the
finite-temperature conductance through a double-quantum-dot (DQD)
system. A smaller dot (``dot 1'') exhibits Kondo physics, while a
larger dot (``dot 2'') is effectively noninteracting and lies near
a transmission resonance. When the dots are connected in parallel
to external leads, and the system is fine-tuned via applied
voltages that determine tunneling barriers and the energies of
individual dot orbitals, a \textit{pseudogap} in the low-energy
effective hybridization between dot 1 and the leads gives rise to
a pair of continuous QPTs between Kondo-screened and non-Kondo
ground states. \cite{Dias:2006} We describe how the system can be
steered into the vicinity of a QPT by monitoring the linear
conductance while changing just two gate voltages.

Experimental detection of QPTs necessarily relies on
finite-temperature manifestations of the underlying $T=0$
transition. We show that the signatures of quantum criticality in
the present DQD system become more pronounced as the temperature
is increased from absolute zero, a trend that contrasts with the
typical behavior near an impurity QPT. \cite{Vojta:1807:2006}
Their temperature dependence also allows these signatures to be
distinguished from other conductance features in the same system.

\textit{Model and conductance calculation.}---Consider a DQD
device in which dot 1 is in an odd-electron-number Coulomb
blockade valley, and dot 2 has a single level near the Fermi level
and is effectively noninteracting. \cite{non-essential} The dots
are coupled to left ($L$) and right ($R$) metallic leads and to
each other via tunneling barriers. This device is described by a
two-impurity Anderson Hamiltonian:
\begin{multline}
\label{Eq:Hamiltonian}
H = \sum_{i,\sigma} \veps_i n_{i\sigma}
  + U_1 n_{1\uparrow} n_{1\downarrow}
  + \sum_{\sigma} \left( \lambda \, a^{\dagger}_{1\sigma}
    a^{\phantom{\dagger}}_{2\sigma} + \mathrm{H.c.} \right) \\
  + \sum_{\ell,\mathbf{k},\sigma} \veps_{\ell\mathbf{k}} \,
    c^{\dagger}_{\ell\mathbf{k}\sigma}
    c^{\phantom{\dagger}}_{\ell\mathbf{k}\sigma}
  + \sum_{i,\ell,\mathbf{k},\sigma} \left( V_{i\ell} \,
     a^{\dagger}_{i\sigma} c^{\phantom{\dagger}}_{\ell\mathbf{k}\sigma}
     + \mathrm{H.c.} \right) ,
\end{multline}
where $a^{\dagger}_{i\sigma}$ creates a spin-$\sigma$ electron in dot $i$
($=1,2$), $n_{i\sigma}=a^{\dagger}_{i \sigma} a^{\phantom{\dagger}}_{i\sigma}$,
and $c^{\dagger}_{\ell\mathbf{k}\sigma}$ creates a spin-$\sigma$ electron of
wave vector $\mathbf{k}$ and energy $\veps_{\ell\mathbf{k}}$ in lead $\ell$
($=L,R$). We assume for simplicity that each lead has a density of states
$\rho(\omega)=\rho_0 \Theta(D-|\omega|)$, symmetric about the Fermi energy
($\omega=0$), and that dot-lead couplings are local. We further assume that
all couplings are real and the device is tuned to left-right symmetry, so that
we can write $V_{i\ell}= V_i / \sqrt{2}$.

The linear conductance at temperature $T$ for this DQD setup can be obtained
from the Landauer formula as:
\begin{gather}
\label{Eq:G}
g(T) = g_0 \int_{-\infty}^{\infty} \! d\omega \,
    \left( -\partial f/\partial\omega \right)
    \left[ -\mathrm{Im} \, \Tm(\omega)\right] , \\
\label{Eq:T}
\Tm(\omega) = 2 \pi \rho_0 \sum_{i,j} V^*_{iL} \, G_{ij}(\omega) \ V_{jR} ,
\end{gather}
where $g_0=2e^2/h$, $f(\omega/T)=[\exp(\omega/T)+1]^{-1}$ is the
Fermi-Dirac function, and all $G_{ij}(\omega)$ in Eq.\
\eqref{Eq:T} are dressed Green's functions, fully taking into
account the electron-electron interactions on dot 1.

The standard equations of motion $\omega \, \langle\!\langle
A;B\rangle\!\rangle_{\omega} - \langle \{ A, B \} \rangle =
\langle\!\langle [ A, H ] ; B \rangle\!\rangle_{\omega} = -
\langle\!\langle A; [ B, H ] \, \rangle\!\rangle_{\omega}$ for the
retarded Green's function $\langle\!\langle
A;B\rangle\!\rangle_{\omega} = -i\int_0^{\infty} dt \, e^{i\omega
t} \langle\{A(t),$ $B(0)\}\rangle$ allow one to re-express
$G_{ij}(\omega)=\langle\!\langle a^{\phantom{\dagger}}_{i\sigma};
a^{\dagger}_{j\sigma} \rangle\!\rangle_{\omega}$ in terms of
$G_{11}$ and the bare Green's function $G^{(0)}_{22}$, which
describes the noninteracting dot 2 in the absence of dot 1. In the
wide-band limit $|\omega|\ll D$, \cite{non-essential} Eq.\
\eqref{Eq:T} becomes
\begin{multline}
\label{Eq:Tfinal}
\Tm(\omega) = \Delta_1 G_{11}(\omega) + 2 \Delta_{12} [G^{(0)}_{22}(\omega)
  \, (\lambda - i\Delta_{12}) \, G_{11}(\omega)] \\
+ \Delta_2 [1 + G^{(0)}_{22}(\omega) \, (\lambda - i\Delta_{12})^2 \,
  G_{11}(\omega)] \, G^{(0)}_{22}(\omega) ,
\end{multline}
where $\Delta_i = \pi \rho_0 V^2_i$, $\Delta_{12} = \pi\rho_0 V_1 V_2$,
and $G^{(0)}_{22}(\omega) = \left(\omega-\veps_2 + i \Delta_2 \right)^{-1}$.

The dot-1 local Green's function $G_{11}(\omega)$ entering Eq.\
\eqref{Eq:Tfinal} can be obtained \cite{Dias:2006} by mapping the
Hamiltonian \eqref{Eq:Hamiltonian} to an effective model of a
single dot connected to the leads via a nonconstant hybridization
function
\begin{equation}
\label{Eq:Delta}
\Delta(\omega) = \pi \rho_2(\omega) \left[\lambda
  + (\omega - \veps_2) \sqrt{\Delta_1/\Delta_2} \right]^2 ,
\end{equation}
with
$\rho_2(\omega)=\Delta_2/\{\pi[(\omega-\veps_2)^2+\Delta_2^2]\}$.
We solve this effective model using the numerical renormalization
group. \cite{Bulla:RMP} At $T>0$, we compute the spectral function
$A_{11}(\omega) = -\pi^{-1}\mathrm{Im}\,G_{11}(\omega)$, and hence
obtain $G'_{11}(\omega) = \mathrm{Re} \, G_{11}(\omega)$ via a
Kramers-Kronig transformation. At $T=0$, where Eq.\ \eqref{Eq:G}
involves only $G_{11}(0)$, it is possible to calculate
$G'_{11}(0)$ directly. All results shown are for $U_1=0.5D$ and
$\Delta_2=0.02D$ with temperatures in units of $T_{K0}=7.0\times
10^{-4}D$, the Kondo temperature in the reference case where dot 2
is decoupled ($\lambda=\Delta_2=0$) and $U_1=-2\veps_1=0.5D$,
$\Delta_1=0.05D$.

To facilitate interpretation of the results, we note that
$-\mathrm{Im}\,\Tm\!(\omega)$ entering Eq.\ \eqref{Eq:G} can be expressed as
\begin{multline}
\label{Eq:ImT}
-\mathrm{Im}\,\Tm\!(\omega) =
   \bigl[ 1 - 2\pi\Delta_2\rho_2(\omega) \bigr] \pi \Delta(\omega)
      A_{11}(\omega) \, + \, \pi \Delta_2 \rho_2(\omega) \\
  + 2 \pi (\omega - \veps_2) \Delta(\omega) \rho_2(\omega) G'_{11}(\omega) .
\end{multline}
The term $\pi \Delta_2 \rho_2(\omega)$ represents bare
transmission through dot 2 in the absence of dot 1, and for
$T\ll\Delta_2$ yields a conductance contribution $g_2\simeq
g_0\Delta_2^2/(\veps_2^2\!+\!\Delta_2^2)$. In most cases of
interest, the term involving $G'_{11}$ turns out to be negligible.
If, as we assume, the dot-1 level is off resonance (i.e.,
$|\veps_1|\gg T,\Delta_1$), then dot 1 appreciably influences $g$
only in the Kondo regime $T\lesssim T_K$ where $A_{11}(\omega)$
exhibits a many-body resonance at the Fermi level; the sign of the
resulting conductance term $g_1$ depends on that of
$1\!-\!2\pi\Delta_2\rho_2(\omega)$ in the range $|\omega|\lesssim
O(T)$ that determines $g(T)$. For $|\veps_2|\gg\Delta_2$
($|\veps_2|\ll\Delta_2$), $g_1$ is positive (negative) at low
temperatures, leading to constructive (destructive) interference
with $g_2$.

\begin{figure}
\includegraphics*[width=0.9\columnwidth]{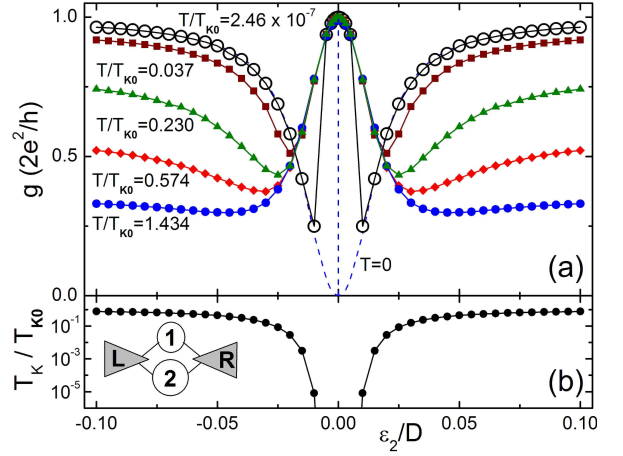}
\caption{\label{fig:Gpar_vs_e2}%
(color online) (a) Conductance $g$ vs $\veps_2$ at six
temperatures for a \textit{parallel} DQD device [inset in (b)]
with $\veps_1=-U_1/2$ and $\Delta_1=0.05D$. (b) Transmission
through dot 1 sets in below the Kondo temperature $T_K$ (defined
as in \protect\onlinecite{Dias:2006}), which vanishes as
$\veps_2\to 0$.}
\end{figure}

\textit{Tuning to the pseudogap regime.}---When the level energy
in dot 2 is set to $\veps_2=\lambda \sqrt{\Delta_2/\Delta_1}$, the
dot-1 effective hybridization [Eq.\ \eqref{Eq:Delta}] vanishes at
the Fermi level as $\Delta(\omega)\propto \omega^2$. The pseudogap
Anderson impurity model, in which $\Delta(\omega) \propto
|\omega|^r$ for $|\omega|\to 0$, exhibits Kondo and non-Kondo
ground states separated by QPTs. Whereas previous theoretical work
\cite{pseudogap-Anderson} has focused on exponents $0< r\leq 1$,
the proposed DQD setup offers a controlled realization of the case
$r=2$, which features a pair of QPTs. For simplicity, we focus in
the remainder of the paper on configurations in which the dots are
connected to the leads purely in parallel, i.e., $\lambda=0$ [see
inset in Fig.\ \ref{fig:Gpar_vs_e2}(b)]. Then Eq.\
\eqref{Eq:Delta} reduces to $\Delta(\omega) =
\Delta_1(\omega-\veps_2)^2 / [(\omega-\veps_2)^2+\Delta_2^2]$.

In order to probe the QPTs, the pseudogap in $\Delta(\omega)$ must
be centered on the Fermi energy. Operationally, this can be
accomplished by tuning $\veps_2$ (via a plunger gate voltage on
dot 2) to reach a maximum of $g$. Figure \ref{fig:Gpar_vs_e2}(a)
illustrates $g$ vs $\veps_2$ at six temperatures, for fixed
$\veps_1=-U_1/2$ and $\Delta_1=0.05D$. For $\lambda=0$,
$\Delta(0)=0$ when dot 2 is exactly in resonance with the leads:
$\veps_2=0$. The choice of $\veps_1 = -U_1/2$ makes $\veps_2=0$ a
point of particle-hole (p-h) symmetry, and ensures that for
$|\veps_2|\ll\Delta_2$ and $T\ll T_K$,
$\pi\Delta(0)A_{11}(0)\simeq 1$; \cite{Vaugier:2007} then, since
$\pi\Delta_2\rho_2(0)\simeq 1$, $g_1$ almost completely cancels
$g_2$. Figure \ref{fig:Gpar_vs_e2}(b) shows that the temperature
range $0\le T\lesssim T_K(\veps_2)$ of the low-conductance regime
shrinks rapidly as $\veps_2\to 0$. For the special case
$\veps_2=0$, the pseudogap in $\Delta(\omega)$ prevents the
formation of a Kondo state (effectively, $T_K=0$), and transport
takes place solely through dot 2. At $T=0$, the resulting
conductance exhibits a discrete jump from $g=0$ for $|\veps_2|\to
0$ to $g=g_0$ for $\veps_2=0$ [dashed line in Fig.\
\ref{fig:Gpar_vs_e2}(a)]. However, this spike broadens at $T>0$
into a smooth peak rising to $g(\veps_2=0)\simeq g_0$.

For a general $\veps_1\ne-U_1/2$, transmission through dot 1 is
still blocked when $\Delta(0)=0$. This leads to an asymmetric peak
in $g(\veps_2)$ at $g(0)\simeq g_0$---a feature that again
broadens with increasing $T$, \cite{in_prep} offering a practical
method for tuning the pseudogap to the Fermi level.

\begin{figure}
\includegraphics*[width=0.98\columnwidth]{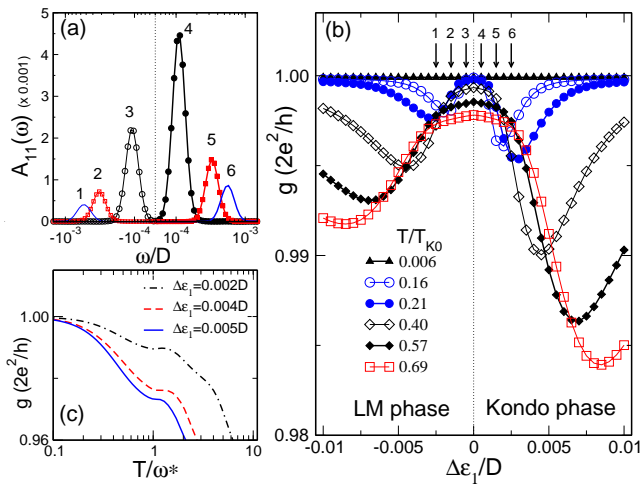}
\vspace{-2ex}
\caption{\label{fig:Gpar_vs_e1}%
(color online) Behavior near the $\veps^+_{1c}$ transition in a
parallel DQD device with $\Delta_1 = 0.05D$ and $\veps_2 = 0$:
(a) Curves 1--6 show the dot-1 spectral function $A_{11}(\omega)$
for the values of $\Delta\veps_1=\veps_1-\veps^+_{1c}$ indicated
by the corresponding arrows in (b). The frequency $\omega^*$
of the quasiparticle peak in $A_{11}(\omega)$ is proportional to
$\Delta\veps_1$.
(b) Conductance $g$ vs $\Delta\veps_1$ at six temperatures.
(c) $g$ vs $T/\omega^*$ at three values of $\Delta\veps_1$ on the
Kondo side of the transition.}
\end{figure}

\textit{Tuning to a QPT.}---With
$\veps_2$ held at zero, the level energy $\veps_1$ can be
varied via a plunger gate voltage on dot 1. A pair of QPTs,
related by p-h duality and located at
$\veps_1=\veps^{\pm}_{1c}=-U_1/2\pm|\Delta\veps_{1c}|$, bound a
local-moment (LM) regime $\veps^-_{1c}<\veps_1<\veps^+_{1c}$ in
which the net spin on dot 1 is unscreened at $T=0$. Close to
either QPT, $A_{11}(\omega)$ contains a quasiparticle peak
centered at $\omega=\omega^*$, where
$\omega^*\propto\veps_1-\veps^{\pm}_{1c}$ [Fig.\
\ref{fig:Gpar_vs_e1}(a)]. The peak sets in below a crossover
temperature $\simeq|\omega^*|$, which on the Kondo side is
proportional to $T_K$. This feature in $A_{11}(\omega)$ leads, via
Eqs.\ \eqref{Eq:G} and \eqref{Eq:ImT}, to a conductance
contribution $g_1<0$ that is greatest in magnitude when
$|\omega^*|\simeq 4T$. Since the dot-2 contribution $g_2\simeq
g_0$ is independent of $\veps_1$, $g$ vs $\veps_1$ isotherms
[e.g., see Fig.\ \ref{fig:Gpar_vs_e1}(b)] show a dip at
$|\veps_1-\veps^{\pm}_{1c}|\propto T$ on either side of a maximum
at $\veps_1=\veps^{\pm}_{1c}$.

It is striking that at $T=0$, the conductance shows no feature as
dot 1 passes through a QPT. At finite temperatures, by contrast,
the DQD device can be tuned to the transition by seeking a local
maximum in $g$ vs $\veps_1$. This maximum has the identifying
characteristics [Fig.\ \ref{fig:Gpar_vs_e1}(b)] that the minima on
either side are equidistant in $\veps_1$ from $\veps^{\pm}_{1c}$,
but the dip in $g$ is roughly twice as deep on the Kondo side,
reflecting the greater weight of the quasiparticle peak in that regime.
For the parameters shown in Fig.\ \ref{fig:Gpar_vs_e1}(b), the
conductance peak becomes more prominent with increasing temperature
up to $T\simeq 3 T_{K0}$, and a peak in $g$ remains discernible
up to the relatively high scale $T \simeq 6 \, T_{K0}$.

The form of $g$ vs $T$ at fixed $\veps_1$ is more complicated since
$g_1$ and $g_2$ can have temperature variations of comparable magnitude.
Figure \ref{fig:Gpar_vs_e1}(c) shows that in the Kondo regime, the peak
in $|g_1(T)|$ contributes a shoulder around $T=|\omega^*|$ to the overall
downward trend dictated by $g_2(T)$. Similar behavior holds in the LM
regime (not shown).

\textit{Differentiating QPT and other conductance features.}---Conductance
peaks similar to those shown in Fig.\ \ref{fig:Gpar_vs_e1}(b) can also
arise, not from proximity to a QPT, but rather from interference between a
conventional (metallic or $r=0$) many-body Kondo resonance on dot 1 and a
noninteracting resonance on dot 2. In experiments, the mapping between the
gate voltages in a real device and parameters of the effective Anderson
model will not be known \textit{a priori}. It is therefore important to be
able to identify unique signatures of a QPT in this system. We show below
that the temperature dependence of the conductance peaks serves this purpose.

\begin{figure}
\includegraphics*[width=0.9\columnwidth]{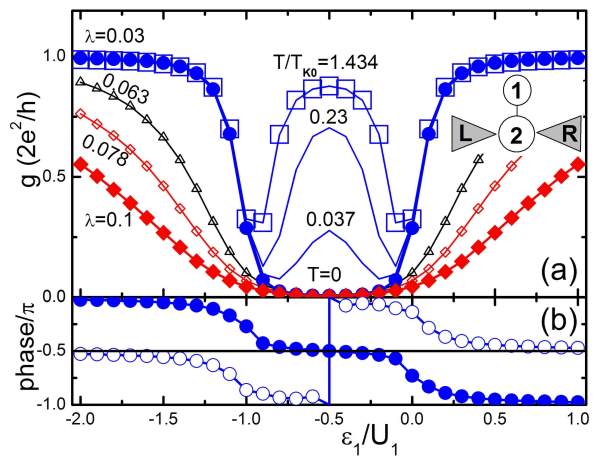}
\caption{\label{fig:Gside_vs_e1}%
(color online) (a) Conductance $g$ vs $\veps_1$ for a \textit{side-dot}
device (inset) with $\veps_2=0$, both for $T=0$ at various $\lambda$ values
and for $\lambda=0.03D$ at the labeled temperatures.
(b) Phase shifts $\eta_{11}$ (filled circles) and $\eta_{22}$ (open circles)
for $\lambda=0.03D$, $T=0$; $g$ vanishes when $\sin\eta_{22}=0$
[Eq.\ \protect\eqref{Eq:Gside-dot}].}
\end{figure}

For simplicity, we consider the ``side-dot'' regime \cite{side-dot_theory}
$\Delta_1=0$ in which dot 1 is connected to the leads only via the
noninteracting dot 2.
In this geometry [inset in Fig.\ \ref{fig:Gside_vs_e1}(a)], the
effective dot-1 hybridization function
$\Delta(\omega)=\pi\lambda^2\rho_2(\omega)$ [from Eq.\
\eqref{Eq:Delta}] is a Lorentzian of width $\Delta_2$ centered at
$\omega = \veps_2$.
Since $\Delta(\omega)$ has no pseudogap, there is no QPT.

Figure \ref{fig:Gside_vs_e1} plots the variation of the conductance with the
position $\veps_1$ of the energy level in the side dot 1 while dot 2 is held
in resonance, i.e., $\veps_2=0$. At $T=0$, the conductance drops to zero as
$\veps_1$ approaches the p-h-symmetric point $\veps_1=-U/2$, independent of
the dot-dot coupling $\lambda$.
This can be understood by noting that for $\Delta_1=0$ and $T=0$,
Eqs.\ \eqref{Eq:G} and \eqref{Eq:T} reduce to
\begin{equation}
\label{Eq:Gside-dot}
g(T=0)=-g_0\Delta_2|G_{22}(0)|\sin\eta_{22} ,
\end{equation}
where $\eta_{ii}=\arg{G_{ii}(0)}$ is the Fermi-energy phase shift
of electrons scattering from dot $i$. Figure
\ref{fig:Gside_vs_e1}(b) shows that in a window about the
p-h-symmetric point, the dot-1 phase shift exhibits a plateau
$\eta_{11}\simeq -\pi/2$ characteristic of the Kondo state.
\cite{Gerland:3710:2000} This additional phase shift of electrons
that scatter from the side dot on their path between the two leads
renormalizes the bare dot-2 phase shift $\eta^{(0)}_{22}=-\pi/2$
to produce an $\eta_{22}$ that jumps from $-\pi$ to 0 at the p-h
point. On moving away from $\veps_1=-U_1/2$, dot 1 gradually
enters its mixed-valence regime, where there is no Kondo resonance
and the $T=0$ conductance rises towards its unitary limit $g=g_0$.
With increasing $\lambda$, the Kondo state in dot 1 becomes more
robust (as evidenced \cite{Dias:2006} by its larger $T_K$),
pushing this upturn in $g$ to larger values of $|\veps_1+U_1/2|$.

Raising the temperature progressively destroys the Kondo resonance
and thereby increases the conductance. For fixed $T>0$, $g$ vs $\veps_1$
reaches a peak at $\veps_1=-U_1/2$, where $T_K$ is smallest and Kondo
scattering is weakest. Figure \ref{fig:Gside_vs_e1}(a) illustrates this
behavior at three temperatures for $\veps_2 = 0$, $\lambda=0.03D$.
The double-dip structure surrounding the peak in $g$ vs $\veps_1$ is
qualitatively similar to the QPT feature in Fig.\ \ref{fig:Gpar_vs_e1}(b).
However, the temperature variation is very different in the two cases.
In Fig.\ \ref{fig:Gpar_vs_e1}(b), the \textit{decrease} with increasing $T$
of the conductance both at the peak and at the minima on either side is
characteristic of the QPT. By contrast, conductance peaks arising for
$\Delta(0) \ne 0$ exhibit an \textit{increase} with $T$ of the extremal $g$
values, as seen in Fig.\ \ref{fig:Gside_vs_e1}(a).

To conclude, we have studied the linear conductance through a class of
quantum-dot devices that can be described by a single Anderson impurity
coupled to a conduction band via a nonconstant hybridization function.
Such devices can be tuned to a quantum phase transition, marked by a
near-unitary peak in the linear conductance that becomes more
pronounced with increasing temperatures. The details of its evolution
with temperature differentiate this conductance signature from similar
features arising from interference effects unrelated to quantum criticality.
Our results demonstrate that these quantum-dot devices offer many
advantages for the controlled experimental investigation of a rich array
of many-body physics.

We thank C.\ Lewenkopf, C.\ B\"{u}sser, and E.\ Vernek for valuable
discussions, and support under NSF-DMR grants 0312939 and 0710540
(Florida), 0336431, 0304314 and 0710581 (Ohio), and 0706020 (Tennessee).


\begin{thebibliography}{18}

\bibitem{Sachdev:book}
  S.\ Sachdev, \textit{Quantum Phase Transitions} (Cambridge University Press,
  Cambridge, U.K., 1999).

\bibitem{Vojta:1807:2006}
  M. Vojta, Phil. Mag. {\bf 86},  1807  (2006).

\bibitem{MIT}
  A.\ Husmann \textit{et al.},
  Science \textbf{274}, 1874 (1996).

\bibitem{HTC}
  S.\ Sachdev, Rev.\ Mod.\ Phys.\ \textbf{75}, 913 (2003).

\bibitem{SIT}
  N.\ Mason and A.\ Kapitulnik,
  Phys.\ Rev. Lett.\ \textbf{82}, 5341 (1999).

\bibitem{qH}
  S.\ L.\ Sondhi, S.\ M.\ Girvin, J.\ P.\ Carini, and D.\ Shahar,
  Rev.\ Mod.\ Phys.\ \textbf{69}, 315 (1997).

\bibitem{Goldhaber-Gordon:156:1998}
  D.\ Goldhaber-Gordon \textit{et al.},
  Nature (London) \textbf{391}, 156 (1998).

\bibitem{multi-dot_expts}
  H.\ Jeong, A.\ M.\ Chang, and M.\ R.\ Melloch,
  Science \textbf{293}, 2221 (2001);
  N.\ J.\ Craig \textit{et al.},
  \textit{ibid.\ } \textbf{304}, 565 (2004);
  J.\ C.\ Chen, A.\ M.\ Chang, and M.\ R.\ Melloch,
  Phys.\ Rev.\ Lett.\ \textbf{92}, 176801 (2004);
  A.\ Fuhrer, T.\ Ihn, K.\ Ensslin, W.\ Wegscheider, and M.\ Bichler,
  \textit{ibid.\ } \textbf{93}, 176803 (2004);
  R.\ Leturcq \textit{et al.},
  \textit{ibid.\ } \textbf{95}, 126603 (2005).


\bibitem{QPTsNRG}
  W.\ Hofstetter and H.\ Schoeller,
  Phys.\ Rev. Lett.\ \textbf{88}, 016803 (2001);
  M.\ Pustilnik, L.\ Borda, L.\ I.\ Glazman, and J.\ von Delft,
  Phys.\ Rev.\ B \textbf{69}, 115316 (2004);
  M.\ R.\ Galpin, D.\ E.\ Logan, and H.\ R.\ Krishnamurthy,
  Phys.\ Rev.\ Lett.\ \textbf{94}, 186406 (2005);
  G.\ Zarand, C.\ H.\ Chung, P Simon, and M.\ Vojta,
  \textit{ibid.\ } \textbf{97}, 166802 (2006);
  R.\ Zitko and J.\ Bonca,
  Phys.\ Rev.\ B \textbf{76}, 241305(R) (2007).

\bibitem{Potok:167:2007}
  R.\ M.\ Potok \textit{et al.},
  Nature (London) \textbf{446}, 167 (2007).

\bibitem{Roch:633:2008}
  N. Roch \textit{et al.}, Nature {\bf 453},  633  (2008).

\bibitem{Dias:2006}
  (a) L.\ G.\ G.\ V.\ Dias da Silva, N.\ P.\ Sandler, K.\ Ingersent,
  and S.\ E.\ Ulloa, Phys.\ Rev.\ Lett.\ \textbf{97}, 096603 (2006);
  (b) \textit{ibid.\ } \textbf{99}, 209702 (2007).

\bibitem{non-essential}
  The assumptions of a noninteracting dot 2 and a wide band simplify the
  analysis, but essentially the same properties arise from the full solution
  of Eq.\ \protect\eqref{Eq:Hamiltonian} for a weakly interacting dot 2,
  as will be discussed elsewhere. \cite{in_prep}

\bibitem{Bulla:RMP}
  R.\ Bulla, T.\ A.\ Costi, and T.\ Pruschke,
  Rev.\ Mod.\ Phys.\ \textbf{80}, 395 (2008).

\bibitem{pseudogap-Anderson}
  D.\ Withoff and E.\ Fradkin,
  Phys.\ Rev.\ Lett.\ \textbf{64}, 1835 (1990);
  C.\ Gonzalez-Buxton and K.\ Ingersent,
  Phys.\ Rev.\ B \textbf{54}, R15614 (1996);
  \textit{ibid.} \textbf{57}, 14254 (1998);
  R.\ Bulla, Th.\ Pruschke, and A.\ C.\ Hewson,
  J.\ Phys.\ Condens.\ Matter \textbf{9}, 10463 (1997);
  M.\ Vojta and R.\ Bulla, Phys.\ Rev.\ B \textbf{65}, 014511 (2001);
  L.\ Fritz and M.\ Vojta, \textit{ibid.\ } \textbf{70}, 214427 (2004).

\bibitem{Vaugier:2007}
  L.\ Vaugier, A.\ A.\ Aligia, and A.\ M.\ Lobos,
  Phys.\ Rev.\ Lett.\ \textbf{99}, 209701 (2007).

\bibitem{in_prep}
  W.\ B.\ Lane, K.\ Ingersent, L.\ G.\ G.\ V.\ Dias da Silva, N.\ P.\ Sandler,
  and S.\ E.\ Ulloa, in preparation.

\bibitem{side-dot_theory}
  K.\ Kang, S.\ Y.\ Cho, J.-J.\ Kim, and S.\-C.\ Shin,
  Phys.\ Rev.\ B \textbf{63}, 113304 (2001);
  V.\ M.\ Apel \textit{et al.},
  Microelectr.\ J.\ \textbf{34}, 729 (2003);
  C.\ A.\ B\"{u}sser \textit{et al.},
  Phys.\ Rev.\ B \textbf{70}, 245303 (2004);
  P.\ S.\ Cornaglia and D.\ R.\ Grempel,
  \textit{ibid.\ } \textbf{71}, 075305 (2005);
  P.\ Simon, J.\ Salomez, and D.\ Feinberg,
  \textit{ibid.\ } \textbf{73}, 205325 (2006);
  R.\ Zitko and J.\ Bonca, \textit{ibid.\ } \textbf{73}, 035332 (2006);
  P.\ A.\ Orellana, G.\ A.\ Lara, and E.\ V.\ Anda, \textit{ibid.\ }
  \textbf{74}, 193315 (2006);
  A.\ C.\ Seridonio, M.\ Yoshida, and L.\ N.\ Oliveira,
  arXiv:cond-mat/0701529 (2007).

\bibitem{Gerland:3710:2000}
  U.\ Gerland, J.\ von Delft, T.\ A.\ Costi, and Y.\ Oreg,
  Phys.\ Rev.\ Lett.\ \textbf{84}, 3710 (2000).
\end{thebibliography}
\end{document}